# Impact of organic spacer cations on the light and thermal stability of two-dimensional lead halide perovskites

*Mahesh Eledath-Changarath[1], Jesús Ortiga-Fibla[1], Andrés F. Gualdrón-Reyes[2,3], Jesús Rodríguez-Romero[2,4], Iván Mora-Seró[2], Ana Cros[1], Núria Garro[1], Rodolfo Canet-Albiach[1], Juan P. Martínez-Pastor[1,6], Andrii Boichuk[1,7], Tetiana Boichuk[1], Juan F. Sánchez-Royo[1,6†], Marie Krecmarova[1*]*

[1]Instituto de Ciencia de Materiales, Universidad de Valencia (ICMUV), 46980 Valencia, Spain.
[2]Institute of Advanced Materials (INAM), Universitat Jaume I, 12006 Castellón, Spain.
[3]Facultad de Ciencias, Instituto de Ciencias Químicas, Isla Teja, Universidad Austral de Chile, 5090000, Valdivia, Chile.
[4]Facultad de Química, Universidad Nacional Autónoma de México, Circuito Exterior s/n, C.U., Coyoacán, 04510 Mexico.
[5]Instituto de Ciencia de Materiales de Madrid (ICMM), CSIC 28049 Madrid, Spain.
[6]MATINÉE: CSIC Associated Unit (ICMM-ICMUV), Universidad de Valencia, Valencia, Spain.
[7] King Danylo University, 76000, Ivano-Frankivsk, Ukraine

[†]Correspondence author: Juan.F.Sanchez@uv.es
[*]Correspondence author: Marie2.Krecmarova@uv.es

**Abstract**

Long-term stability remains a critical challenge limiting the photovoltaic and optoelectronic applications of two-dimensional lead halide perovskites. In these materials, the choice of the organic spacer plays a key role not only in enhancing environmental stability, but also in determining their structural characteristics and carrier dynamics. Here, we present an environmental photo- and thermal-stability analysis of $BA_2PbI_4$ and $PEA_2PbI_4$ two-dimensional lead halide perovskites by means of confocal micro-Raman and X-ray photoelectron spectroscopy techniques. We have investigated two different organic spacers, n-butylammonium (BA) and 2-phenylethylammonium (PEA), which provide comparable room temperature luminescence in freshly exfoliated perovskite nanoflakes and degrade similarly through the removal of organic components and fast crystal phase decomposition into $PbI_2$ and $Pb^0$. However, we observed that the PEA organic cation significantly improves crystal morphology and suppresses lattice defects, resulting in slower degradation and enhanced resistance to light and thermal stress under ambient conditions. Our findings indicate that structural and chemical engineering, along with surface passivation and encapsulation strategies, are essential for achieving long-term stability in two-dimensional halide perovskite devices. Nevertheless, optimizing the balance between stability and device performance remains a critical consideration.

## 1. Introduction

In recent years, two-dimensional (2D) Ruddlesden-Popper perovskites (2DPVKs) gained increased attention due to their tunable optoelectronic properties, better environmental stability, and large exciton binding energy compared to conventional 3D organic-inorganic halide perovskites [1]. The general formula of such 2D perovskite is $(RNH_3)_2 A_{n-1} M_n X_{3n+1}$, where R is a large aliphatic or aromatic alkyl ammonium spacer cation, A is a monovalent organic cation, M is a divalent metal cation, and X is a halide anion [2]. They form a spontaneous quantum well structure which consists of n layers of $[MX_6]^{4-}$ octahedral units that are sandwiched between two layers of large organic spacer cations[2]. Collaterally, the hydrophobic nature of the organic spacer prevents the direct exposure of metal halide octahedral sheets to the air moisture and thus enhances the stability of the 2DPVKs [1]. Excitons, quasi-particles governing the light-matter interaction in semiconductors, are confined in 2DPVKs in the inorganic layers [2–4]. This fact, together with the low dielectric screening of the surrounding organic spacers, leads to a considerable increase in the exciton binding energy [2,3,5]. Actually, the exciton binding energy can be tuned by changing the number of inorganic sheets (n) and by choosing different organic spacers [6]. In contrast to their 3D counterparts, these properties make 2DPVKs an excellent candidate for optoelectronic applications, including photodetectors, lasers, light-emitting diodes, and solar cells.

One of the challenges with perovskite materials, both 3D and 2D, is their liability to degradation when exposed to ambient air [5–9], moisture [10,11], heat [12–14], or light [1,6,13,15,16]which compromise the performance and stability of devices incorporating these materials. Although 2DPVKs present enhanced stability as compared to their 3D counterparts, they are still prone to degrade when exposed to environmental factors over time. Therefore, understanding the underlying mechanisms of degradation and developing appropriate mitigation strategies is paramount for making perovskite-based devices more durable and practical for real-world applications.

The degradation process of perovskites comprises a sequence of chemical reactions.

$$2DPVKs \rightarrow PbI_{2\ solid} + Organic_{\ gas} + HI_{\ gas} \quad (1)$$

$$PbI_{2\ solid} \rightarrow Pb^{0}_{\ solid} + I_{2\ gas} \quad (2)$$

At the early stages, $PbI_2$ is produced, accompanied by the liberation of organic cations from the lattice and the emission of HI gas (equation 1) [1]. Subsequently, $PbI_2$ undergoes decomposition into metallic $Pb^0$ (equation 2) [1,5], which might further break down into lead oxides and hydroxides upon prolonged exposure to oxygen and moisture [1,17,18]. The migration of ions, which results in the intrinsic vacancies and defects in the lattice, is reported to be the primary cause of perovskite degradation [19]. Yuan and co-workers reported the degradation of the methylammonium lead iodide perovskite structures through the ion migration driven by light [20]. Similarly, deQuilettes et. al studied the photo-induced halide ions migration and correlated it with changes in the photoluminescence (PL) [21]. Fang et al. investigated the light-induced degradation of the nanoflakes of 2D perovskite $(PEA)_2PbI_4$ [1]. In our previous work, we demonstrated that light is the most critical external factor accelerating 2DPVKs degradation processes in ambient air [6,16].

The stability of these materials can be modulated through strategies such as surface passivation and encapsulation[1,7,22]. Alternatively, stability can be enhanced by employing structural and compositional molecular engineering approaches [8,9,23]. In fact, modifying the composition of 2DPVKs with carefully selected organic spacer molecules has been shown to improve phase stability, enhance crystal quality, and reduce lattice defects [23–26]. These results already indicate that compositional and morphological engineering strategies, followed by systematic environmental stability and performance analysis, are essential to develop high-performance and durable 2DPVK-based devices [23].

In this work, we systematically study the photo- and thermal-stability of mechanically exfoliated $(BA)_2PbI_4$ and $(PEA)_2PbI_4$ lead halide 2DPVKs flakes under ambient air conditions. Both freshly exfoliated samples have high crystallinity and phase purity as confirmed by the powder X-ray diffraction (XRD) analysis. Specifically, we analyse the influence of two prominent organic spacer molecules: aromatic and rigid 2-phenylethylammonium (PEA) and aliphatic and flexible n-butylammonium (BA) amine ions. The type of organic spacer molecule significantly impacts the environmental stability of the exfoliated flakes [27–31]. A comprehensive set of characterization techniques, including micro-Raman spectroscopy and X-ray photoelectron spectroscopy, was employed to investigate the stability of these materials. Degradation was corroborated by the concurrent observations of the flakes under an optical microscope, where visible changes were observed over time. To clarify the effect of BA and PEA organic spacer cations as a quantum well passivation layer, first of all, we study photostability in atmospheric air. Raman spectroscopy revealed a gradual crystal decomposition accompanied by a decline in PL intensity under prolonged laser illumination, providing clear evidence of photo-induced degradation of the perovskite crystals. Subsequent thermal stability assessments, conducted through XPS and Raman spectroscopy, highlighted significant differences between fresh flakes and those subjected to thermal treatments, conclusively demonstrating the thermal instability of the exfoliated $(BA)_2PbI_4$ and $(PEA)_2PbI_4$ nanoflakes. However, we found that the PEA organic cation significantly enhances the photo- and thermal-stability of 2DPVKs, owing to its rigid molecular structure. This rigid structure contribute to improved crystal morphology, suppression of lattice defects, and increased thermal conductivity, collectively leading to enhanced material stability [24,29,30]. Our findings are important to solve stability limitations of 2DPVKs and their implementation into long-term stable photovoltaic and optoelectronic devices.

## 2. Results and Discussion

### 2.1 Synthesis and characterization

Two types of 2D Ruddlesden-Popper perovskite single crystals with the lowest quantum well thickness (n=1) were synthesised (see materials and methods for detailed information). Figure 1(a)-1(c) presents the crystal structure of $PEA_2PbI_4$ - $(C_8H_{12}N)_2PbI_4$ and $BA_2PbI_4$ – $(C_4H_{12}N)_2PbI_4$ perovskites. The crystal lattice of $PEA_2PbI_4$ and $BA_2PbI_4$ consists of a single layer (n=1) of $[PbI_6]^{4-}$ inorganic octahedral unit that is sandwiched by organic spacer cations $PEA^+$ (hydrophobic, longer length and rigid molecule) and $BA^+$ (shorter length and softer molecule), respectively, forming 2D quantum well structures. The nature of organic cations plays a critical role in determining the crystal structure, including the crystal symmetry and defect concentration [32]. These structural variations, in turn, have a pronounced impact on the optical properties, charge transport, and environmental stability of 2DPVKs [24,29,30]. Synthesized crystals can be readily exfoliated down to a monolayer, due to the relatively weak van der Waals interlayer coupling. In the present work, perovskite flakes were obtained by micromechanical exfoliation of single crystals of both perovskites and deposited onto $SiO_2$/Si substrates (see materials and methods for details). Optical images of freshly exfoliated of $PEA_2PbI_4$ and $BA_2PbI_4$ flakes are shown in Figure 1d and 1e, respectively. The exfoliated single-crystal flakes typically exhibit lateral dimensions ranging from tens to hundreds of micrometres and thicknesses on the order of tens of nanometres as confirmed by atomic force microscopy (AFM) and scanning electron microscopy (SEM) measurements shown in Supplementary material (SI), Figure S1 (AFM) and Figure S2 (SEM). Exfoliation suppresses the formation of unwanted hybrid phase and yields flakes with atomically flat surfaces, with the crystal planes oriented parallel to the substrate. The exciton binding energies of $PEA_2PbI_4$ and $BA_2PbI_4$ are reported to be 0.23 eV [33] and 0.19 eV [34], respectively with band gap around ~2.55 eV [33,34]. Excitons are less mobile in $(BA)_2PbI_4$ (difusivity = 0.013±0.002 $cm^2$/s) thanin $(PEA)_2PbI_4$ (difusivity = 0.192 ± 0.013 $cm^2$) [35]. Figure 2a presents the typical room-temperature PL spectra obtained from freshly exfoliated flakes of $BA_2PbI_4$ and $PEA_2PbI_4$ recorded by using a 488 nm laser excitation source. The spectra feature a prominent high-intensity peak at 2.35 ± 0.005 eV for both perovskites (Figure S3 in SI), attributed to the recombination of free excitons, and allows us to conclude that freshly exfoliated perovskites exhibit similar optical properties and performance. Yet, as we demonstrate in the following section, $PEA_2PbI_4$ and $BA_2PbI_4$ crystals differ markedly in their environmental stability, a difference that may be decisive for their integration into optoelectronic devices.

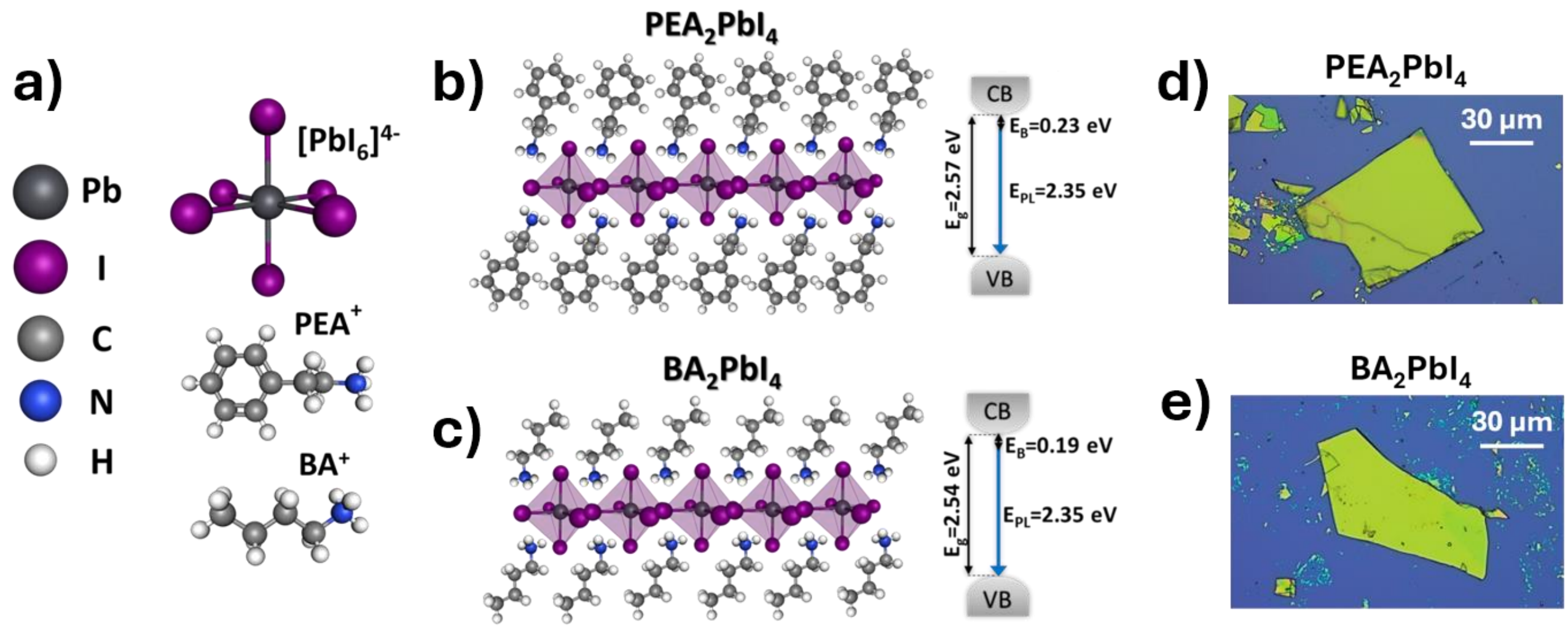

***Figure 1.*** *(a) Composition of 2D lead halide perovskites, showing the inorganic* $(PbI_6)^{4-}$ *octahedral units, and the organic spacer cations Phenethylammonium* $PEA^+$ *or Butylammonium* $BA^+$*. Crystal structure of a monolayer of (b)* $PEA_2PbI_4$ *and (c)* $BA_2PbI_4$ *crystals. The inset shows its energy level diagrams, with valence band (VB), conduction band (CB), electronic bandgap (*$E_g$*), free exciton optical transition (*$E_{PL}$*), and the exciton binding energy (*$E_b$*), including their values obtained from references* [33,34]*. Optical images of freshly exfoliated flakes of (d)* $PEA_2PbI_4$ *and (e)* $BA_2PbI_4$ *crystals.*

To probe crystallinity and phase purity of freshly synthesised samples we employed the powder X-ray diffraction (XRD) analysis. The XRD patterns of $PEA_2PbI_4$ and $BA_2PbI_4$ (Figure 2b) show excellent agreement with the simulated patterns obtained from single-crystal data [36,37], confirming the formation of the desired crystalline phase and the absence of secondary phases. Furthermore, a pronounced preferential orientation is observed, evidenced by the presence of characteristic highly oriented (001) crystallographic family reflections occurring at 5.48° (002), 10.88° (004), 16.32° (006), 21.8° (008), 27.32° (0011), and 32.94° (0012) for $PEA_2PbI_4$ [38,39] and at 6.48° (002), 12.88° (004), 19.36° (006), 25.88° (008), and 32.48° (0011) for $BA_2PbI_4$ [39,40]. This behavior is typical of two-dimensional perovskites, in which the orientation of the crystalline layers plays a fundamental role in their structural and functional properties.

Micro-Raman spectroscopy is a powerful technique to probe the structural properties as well as compositional chemical variations of the perovskites on the micrometre scale. Monitoring the evolution of Raman vibrational modes during degradation processes indicate structural properties such as lattice distortions, organic cation volatilization, and crystal destruction [41,42]. freshly exfoliated and treated $(PEA)_2PbI_4$ and $(BA)_2PbI_4$ thin flakes were investigated using confocal micro-Raman spectroscopy and optical microscopy imaging. First of all, we investigate the chemical composition and crystal structure of both types of freshly exfoliated samples to compare their quality. Room temperature Raman spectra of freshly exfoliated $(PEA)_2PbI_4$ and $(BA)_2PbI_4$, measured in air (obtained using 785 nm laser excitation) are depicted in Figs 3c-d, respectively. The Raman spectra were deconvoluted by Gaussian fitting into 5 components (peaks 1-5) corresponding to specific phonon vibration modes of the inorganic lead octahedra lattice coupled with organic cations. Specifically, Peak 1 ($\sim 75 \pm 2$ $cm^{-1}$) is assigned to in-plane I–Pb–I octahedral bending, Peak 2 ($\sim 86 \pm 3$ $cm^{-1}$) to combination of out-of-plane Pb–I bending and octahedral tilting, Peak 3 ($\sim 95 \pm 3$ $cm^{-1}$) to symmetric Pb–I stretching, Peak 4 ($\sim 107 \pm 4$ $cm^{-1}$) to asymmetric Pb–I stretching (typicaly infrared active and Raman inactive) and $PbI_2$ phase degradation indicator, and Peak 5 ($\sim 126 \pm 2$ $cm^{-1}$) to organic-inorganic hydrogen-bonding [43–52]. A broader feature ($\sim 150 \pm 10$ $cm^{-1}$) is attributed to the interaction between the organic cation and the inorganic framework.

For both samples, there are visible 4 Raman active vibration modes (Peak 1, 2, 3 and 5). Note that these Raman modes are also thickness dependent [48], showing a frequency blue shift (up to several $cm^{-1}$) from monolayer to bulk, due to the suppression of atomic vibrations by interlayer vdW forces [48]. Occurrence of infrared active mode (Peak 4) in freshly exfoliated $(BA)_2PbI_4$ samples might get Raman active due to their softer crystal structure in comparison to the rigid crystal structure of $(PEA)_2PbI_4$ [45,51,53]. Because of both freshly exfoliated samples were monitored by Raman spectroscopy using the safe sub-bandgap (784 nm) excitation laser energy and XRD analysis (Figure 2b) confirmed their high crystallinity and phase purity, we assume that presence of peak 4 in fresh samples is not assigned to crystal degradation to $PbI_2$ phase.

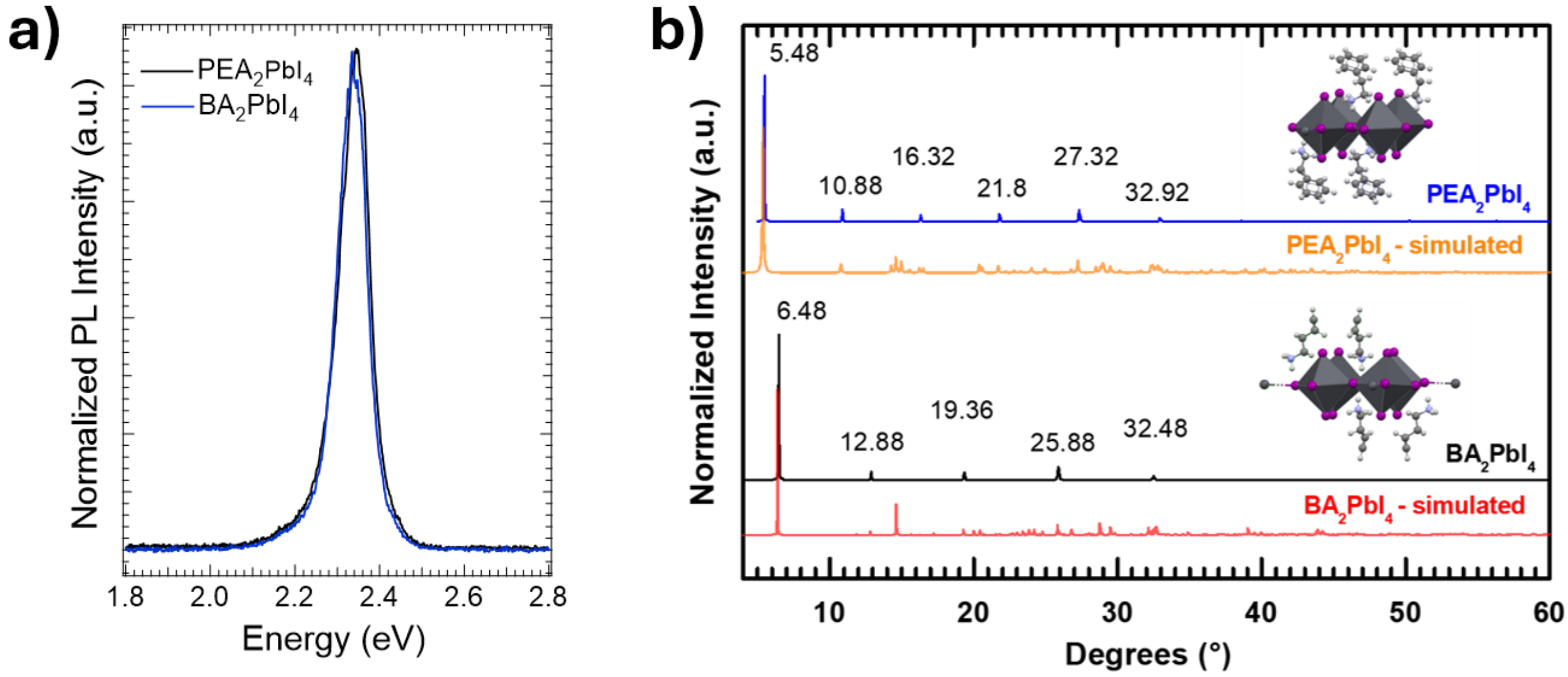

***Figure 2.*** *(a) µ-PL spectrum of $PEA_2PbI_4$ and $BA_2PbI_4$. (b) Powder X-ray diffraction (XRD) patterns of $PEA_2PbI_4$ and $BA_2PbI_4$. Insight depicts the optimized structures of $PEA_2PbI_4$ and $BA_2PbI_4$ double octahedra molecules, with colored atoms as follows: hydrogen (white), carbon (grey), nitrogen (blue), iodine (purple), and lead (dark grey).*

**2.2 Photostability Analysis**

Next, we focused on the analysis of the photostability in ambient air conditions, performed by Raman spectroscopy. In principle, light-induced degradation processes depend on the photon source energy, intensity, illumination time, and especially on the surrounding environmental conditions. Prolonged light illumination can accelerate ion migration and defect accumulation [54]. High impact of light on the crystal structure degradation of freshly exfoliated $PEA_2PbI_4$ and $BA_2PbI_4$ samples exposed to ambient air over the period of 3 weeks with absence and presence of ambient light is confirmed in Figure S4 in SI. Degradation processes are faster in atmospheric air conditions, whereas they are significantly suppressed in vacuum[15] and dark[16]. Higher photon energy and intensity lead to faster crystal decomposition, accelerated in air due to the presence of humidity and oxygen [6,15,55]. For example, illumination with a light source having a higher energy than the 2DPVKs bandgap ~2.56 eV (486 nm, Figure 1) under dry air would cause degradation triggered by photogenerated electrons and superoxide formation [23]. On the other hand, light sources with higher energy than 2.72 eV (~456 nm), such as blue or UV light, would lead to the faster and irreversible collapse of the crystal structure of 2DPVKs, resulting in organic molecule deprotonation, ion migration, defect formation, and phase segregation.

Photodegradation was examined under ambient atmospheric conditions by irradiating the samples with a high-power green laser (λ = 532 nm) fosused on the 2DPVKs nanoflake surface with laser spot diameter ≈3.4 µm , varying both power and illumination time (see Materials and Methods section for more information). To monitor the degradation effect, Raman spectra were recorded immediatelly after each illumination period at the same sample location, but using low energy (λ =785 nm) excitation. The use of a laser with a wavelength of 785 nm with excitation energy lower than optical banggap of investigated samples minimizes their photodegradation and suppresses photoluminescence background. excitation energy effectively minimized uncontrolled photodegradation of the perovskite crystals caused during the Raman measurements, as no observable degradation occurred under 785 nm laser excitation, even at the highest laser power. Raman spectroscopy examination was complemented by optical microscopy imaging [56,57] during photodegradation. Optical images of $(PEA)_2PbI_4$ and $(BA)_2PbI_4$ freshly exfoliated flakes and photodegraded flakes (120 s exposure to 532 nm laser in air) with low (68 µW) and high (4.5 mW) laser powers are depicted in Figure 3 a-b, respectively. Optical micrographs indicate similar degradation in both samples. The laser induces visible holes in both crystals, with higher power correlating with larger damage characterized by larger spot diameters, deeper ablation, and noticeable degradation across the entire crystal surface. Raman spectra corresponding to the optical images (Figures 3a and 3b) of pristine and photodegraded $(PEA)_2PbI_4$ and $(BA)_2PbI_4$ samples are shown in Figures 3c and 3d, respectively. We observe that photodegradation processes strongly depend on the illumination conditions: low laser power (68 µW) does not produce a significant change in the Raman spectra in comparison to the freshly exfoliated sample; higher laser power (4.5 mW) leads to accelerated ion migration and crystal decomposition into the $PbI_2$ accompanied by evaporation of volatile organic cations, consistent with our previous work[6]. Noticeable shifts in Raman mode frequencies and changes in peak intensities were observed in the spectra. These variations can be attributed to alterations in the crystal lattice structure induced by photodegradation. The shift of Raman modes (peaks 1-3) might reflect a flake thickness reduction caused by vulnerability of organic and halide ions, accompanied by crystal decomposition into $PbI_2$ [47,48]. In addition, vacancy defects or absorbed humidity might deform the crystal structure, causing Raman peak shifts [58]. Finally, the frequency shift might also be affected by sample heating induced by high-power laser illumination [59]. Furthermore, monitoring changes in the intensity of Raman peaks may serve as a crucial indicator of chemical and structural transformations occurring during the degradation process.

To understand the mechanisms of the studied laser light-induced degradation processes on both samples, Raman spectra were deconvoluted into 5 components by Gaussian fitting. For both samples, $PEA_2PbI_4$ (Figure 3c) and $BA_2PbI_4$ (Figure 3d), the integrated intensity of peaks 2-3 did not change significantly during photodegradation. On the other hand, monitoring the integrated intensity variations of the peak 4 and 5 vibrational Raman modes provides valuable insight into the degradation mechanisms. Relative contribution of each deconvoluted component to the total spectrum for $PEA_2PbI_4$ (Figure 3c) and $BA_2PbI_4$ (Figure 3d) was calculated to reflect their chemical and structural phase state variation during photodegradation. Figure 3e-h shows changes in the relative contributions of individual components during the laser treatment as a function of laser illumination power (68 µW and 4.5 mW) and laser exposure time (up to 120 s) to compare the degradation behavior of $PEA_2PbI_4$ and $BA_2PbI_4$ samples. $PEA_2PbI_4$ (Figure 3e,g) shows a noticeably higher proportion of peak 5 (organic-inorganic hydrogen-bonding) and a lower proportion of peak 4 (asymmetric Pb–I stretching and $PbI_2$ phase indicator) .whereas in $BA_2PbI_4$ (Figure 3f,i) the relative contribution of Peak 4 is higher, confirming higher resistance of $PEA_2PbI_4$ samples against photodegradationof crystal structure. Relative contribution of individual components doesn't significantly change during the laser treatment using lower laser power illumination for both samples.

On the contrary, their relative contribution notably changes with the high laser power treatment. For both samples, the same trend was observed during the degradation, showing a decrease of peak 5 contribution, attributed to the evaporation of the organic cation, and an increase of peak 4 contribution associated to crystal defects, iodine migrations, among other and formation of $PbI_2$ phase. Consistently for both samples, there is no noticeable sample degradation by lowest laser power illumination, but they degrade faster at high power illumination in air and with increasing treatment duration. By comparing photostability of both samples, it can be concluded that $PEA_2PbI_4$ demonstrated greater resistance to high-power laser illumination under ambient air conditions, with degradation becoming apparent only after 90 seconds of exposure (Figure 3g). In contrast, $BA_2PbI_4$ exhibited signs of degradation already after 60 seconds of laser irradiation(Figure3h).

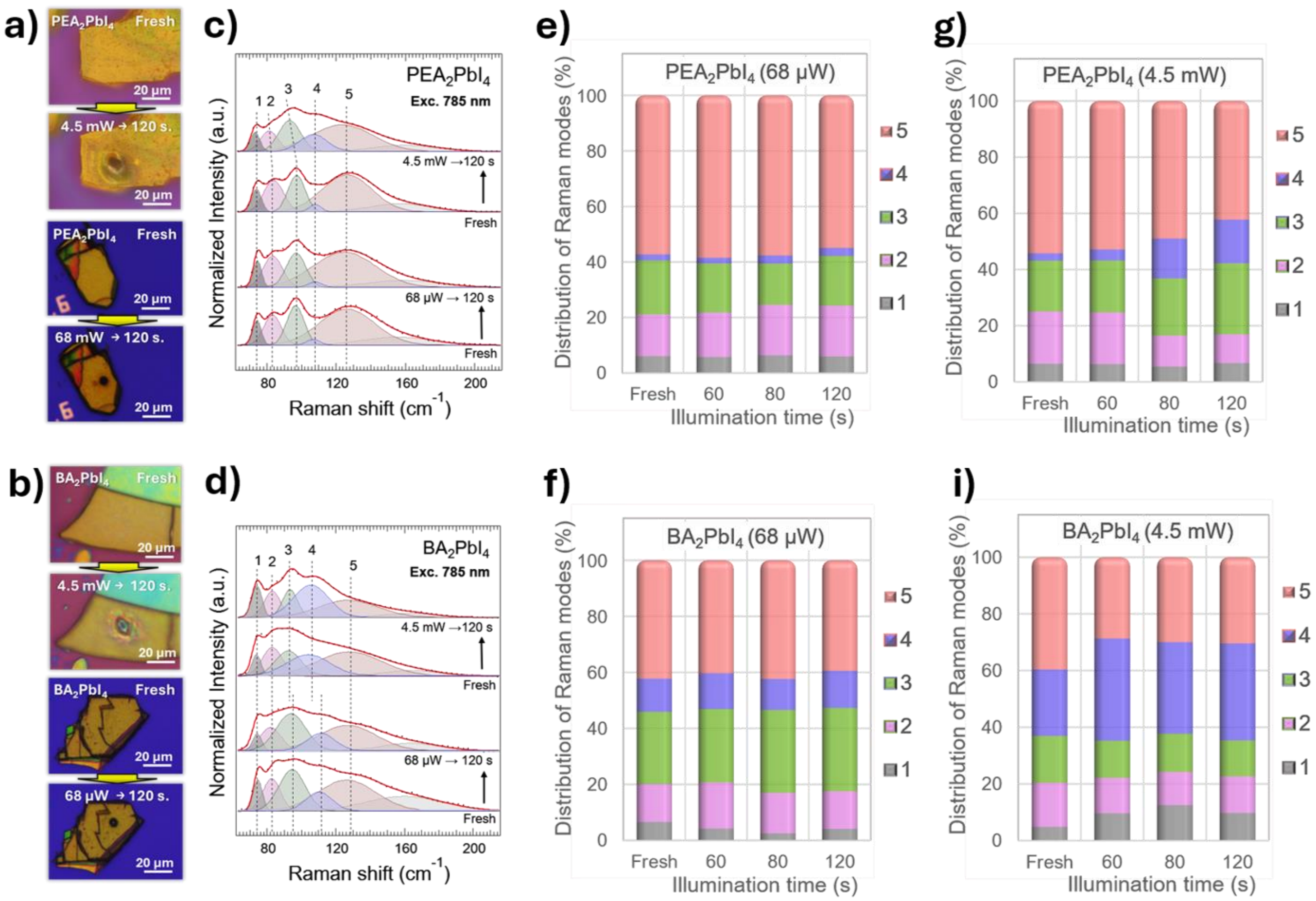

***Figure 3.*** *Optical images of freshly exfoliated (a) $PEA_2PbI_4$ and (b) $BA_2PbI_4$ flakes and after laser beam illumination (532 nm , 120 s duration in air) with low (68 µW) and high (4.5 mW) power. Micro-Raman spectra were recorded immediately after this treatment by using an infrared laser as excitation source, avoiding further photodegradation (785 nm photon energy, 6.4 mW, 90 s). Micro-Raman spectra of freshly exfoliated (c) $PEA_2PbI_4$ and (d) $BA_2PbI_4$ flakes, before and after laser treatment with low (68 µW) and high (4.5 mW) laser power. Evolution of the relative contribution of each deconvoluted Raman vibrational mode to the total spectrum (c) for $PEA_2PbI_4$ as function of laser illumination power of (e) 68 µW and (f) 4.5 mW and duration of the treatment. Evolution of the relative contribution of each deconvoluted Raman vibrational mode to the total spectrum (d) for $BA_2PbI_4$ as function of laser illumination power of (g) 68 µW and (h) 4.5 mW and duration of the laser treatment.*

To correlate structural and chemical changes during the photodegradation processes, PL was examined. PL background was recorded by micro-Raman spectroscopy by using short exposure times and a low-power laser beam (532 nm wavelength, 7.4 µW, 0.15 s) to avoid sample degradation. The PL background is compared for freshly exfoliated $PEA_2PbI_4$ (Figure S5a in Supplementary material) and $BA_2PbI_4$ (Figure S5b in Supplementary material) flakes, before and after photodegradation with low (68 µW) and high (4.5 mW) laser power. The conditions for photodegradation were identical to those of the Raman study described above, but with a duration of 120 s. Consistently for both samples, the overall PL intensity decreases somewhat after low laser power illumination; on the other hand, the intensity decrease is very pronounced after high power illumination.

Photodegradation of 2DPVKs has been previously reported [1,1,11,60] , induced by laser illumination and assigned to degradation processes initiated by volatile organic cation and iodine anion evaporation, followed by decomposition into $Pb^0$ and $PbI_2$ [1,61–63]. Laser irradiation of 2D materials under vacuum conditions can increase the local tempearture up to 213 ◦C at the irradiated spot [64]. High-power laser illumination in air may induce a considerable temperature rise of the crystal surface [1], accelerating ion migration, promoting evaporation of the organic cations, and triggering phase segregation within the crystal[23,65]. We have previously found that photodegradation processes are also dependent on the organic cations [32], where $(PEA)_2PbI_4$ crystals show a higher stability than $(BA)_2PbI_4$ ones [19,61,66,67]. This conclusion is further supported by the Raman scattering analysis performed above. Light-induced degradation in ambient air is accelerated when the photon energy exceeds the material's bandgap, resulting in localized sample heating and surface degradation that initiates at the laser irradiation spot followed by diffusion of excited electrons and a large area lattice phonon transport [65]. The following section addresses heat-induced degradation of the samples under ambient conditions, as a counterpart to the photodegradation processes discussed above. This will allow us to further elucidate the underlying degradation mechanisms.

**2.3 Thermal stability Analysis**

We employed temperature-dependent micro-Raman spectroscopy and optical microscopy techniques to study the thermal stability of the samples. The effect of thermal degradation under ambient atmospheric conditions was investigated as a function of increasing sample temperature, at 25 °C, 40 °C, 80 °C, and 120 °C. Once the desired temperature was reached, the system was stabilized for five minutes to ensure thermal equilibrium conditions. Then, Raman spectra were acquired using low-energy laser excitation (λ = 785 nm) to minimize laser-induced degradation (see Materials and Methods for more information). Optical images of the freshly exfoliated $(PEA)_2PbI_4$ and $(BA)_2PbI_4$ flakes were also recorded at each temperature and are shown in Figure 4a,b for room temperature (25°C), 80°C and 120°C for comparison, since visible changes started at 80°C. The images show sample color variation, attributed to changes in the thickness of the perovskite flakes. Indeed, due to light interference effects, the color of many 2D materials strongly correlates with thickness [56,57,68]. In our particular case, the color change might be associated more preferiably with the change in chemical composition than with the thickness of the flake, since at 120 °C remains is practically $PbI_2$. Optical changes are similar in both 2DPVKs samples, resulting in a similar degradation, with large volume loss.

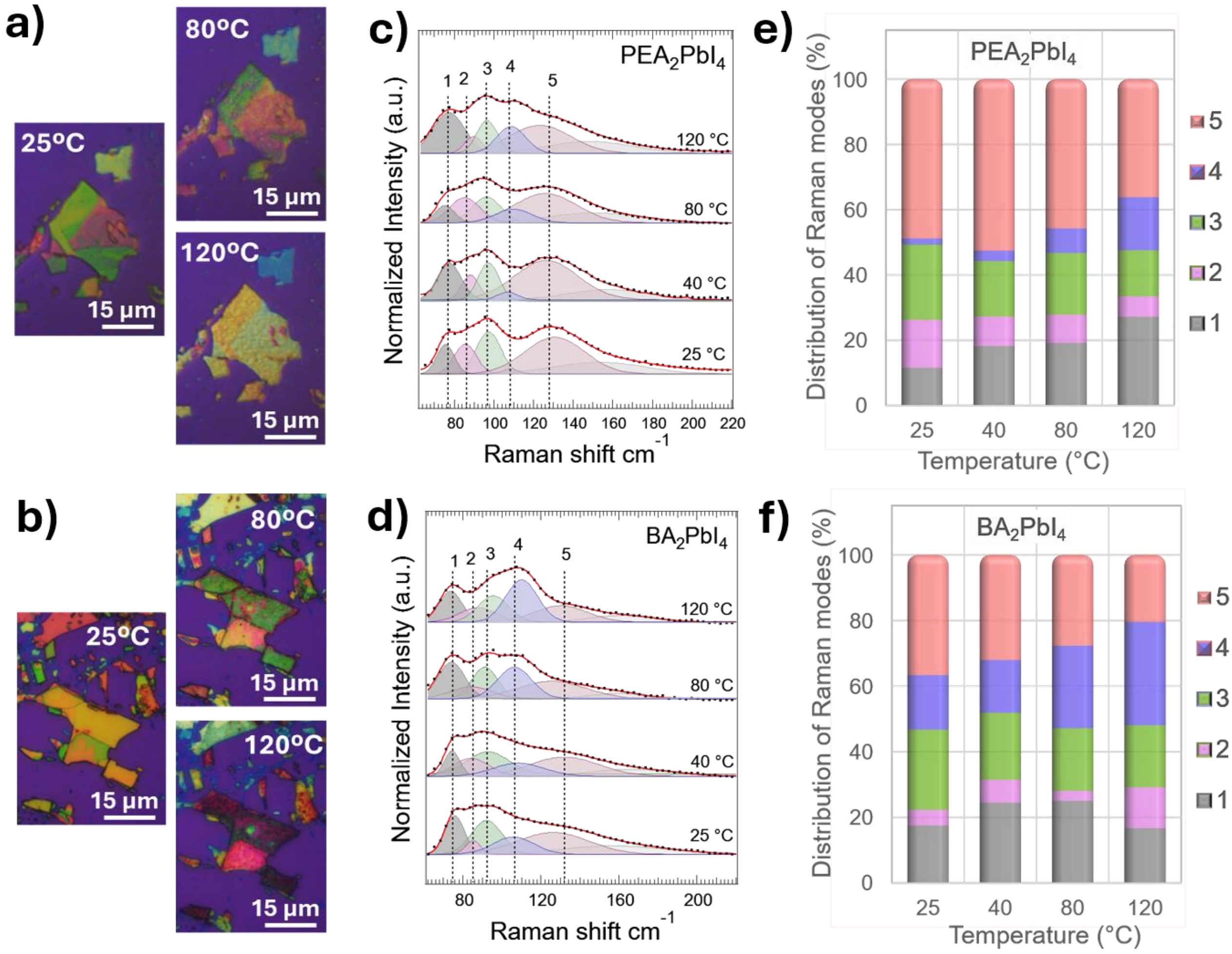

***Figure 4.*** *Optical images of freshly exfoliated (a) $PEA_2PbI_4$ and (b) $BA_2PbI_4$ flakes at room temperature (25 $^oC$), 80, and 120 $^oC$ in air. Micro-Raman spectra (785 nmExcitation, 6.4 mW, 90 s) of freshly exfoliated and thermally treated (c) $PEA_2PbI_4$ and (d) $BA_2PbI_4$ flakes (25 $^oC$, 40 $^oC$, 80 $^oC$ and 120 $^oC$) in air, which were deconvoluted to the 5 individual components. Evolution of the relative contribution of each deconvoluted Raman vibrational mode to the total spectrum (c-d) as function of heating temperature for (e) $PEA_2PbI_4$ and (f) $BA_2PbI_4$ samples.*

Raman spectra corresponding to the optical images (Figure 4a and 4b) of pristine and thermally treated $(PEA)_2PbI_4$ and $(BA)_2PbI_4$ flakes in air are shown in Figure 4c and 4d, respectively. As described above, the spectra can be deconvoluted into the 5 components (Peaks 1-5). Upon increasing the temperature, a noticeable quenching of the peak associated with the vibration of organic cations is observed, indicating their removal from the perovskite lattice. Consistently with the photodegradation experiment, the intensity of peak 4 related vibration mode increases and peak 5 related vibration mode decreases with heating temperature, indicating crystal degradation (see section 2.2). The temperature-dependent Raman analysis, supported with changes in the optical micrographs, confirmed that a significant sample volume loss and structural collapse occurred with increasing temperature. This is further confirmed by the quantitative analysis of changes in relative contribution of each deconvoluted component to the total spectrum (Figure 4c,d) as shown in Figure 4e,f. Hence, with increasing temperature, both samples suffer similar degradation processes. However, comparing thermal stability of both samples, $PEA_2PbI_4$ shows a higher resistance against thermal decomposition than $BA_2PbI_4$. Material containing PEA, despite eventually degrading, exhibits a broader temperature resistance window, maintaining much of its chemical and structural integrity even at 80 °C, where as BA containing sample shows a noticeable degradation already at 40 °C. Additionally, it's worth noting that the aliphatic ammonium salt appears to be considerably less stable than secondary aliphatic cations. A previous study conducted by us evaluated the effect of the dipropylammonium cation [69]. In summary, we propose that the search for more stable materials should prioritize aromatic and secondary aliphatic cations, and even tertiary ones—provided they form the perovskite-type phase.

Thermal stability of 2DPVKs can be enhanced by structural compositional modification engineering [70,71]. 2DPVKs present ultra-low thermal conductivities ranging from 0.10 to 0.19 $W \cdot m^{-1} \cdot K^{-1}$, much lower than their 3DPVKs counterpart [72], depending mainly on the organic cation structure[72,73]. Thermal conductivities of $BA_2PbI_4$ and $PEA_2PbI_4$ are 0.10 and 0.15 $W \cdot m^{-1} \cdot K^{-1}$ [74], respectively. Basically, high temperatures promote 2DPVKs chain fragmentation with further volatilization. Increasing thermal conductivity with an increasing organic chain length and parallel surface orientation related to the substrate induces efficient thermal transport along the surface of the inorganic layered material [73]. Even 2DPVKS having the same chemical compositions, but varying in organic spacer layer orientation, present significant changes in thermal transport across their structure [72]. It follows that the organic spacer molecule orientation has a significant influence on the 2DPVKs thermal conductivity. We assume that a higher thermal conductivity of $(PEA)_2PbI_4$ together with a strong chemical bonding between the $PEA^+$ and $PbI_6^{4-}$ octahedra ensure a crystal quality enhancement, leading to an enhancement in photo and thermostability of $(PEA)_2PbI_4$ 2DPVK [29]. In summary, engineered 2DPVKs with targeted chemical and physical properties are promising for future optoelectronic and thermoelectric applications with excellent device stabilities [72].

In addition to the enhanced thermostability of $(PEA)_2PbI_4$ 2DPVK observed via Raman spectroscopy, X-ray photoelectron spectroscopy (XPS) was employed to investigate the differences in degradation mechanisms of both samples by examining their final surface chemical compositions.XPS stands out as a powerful analytical technique for probing the chemical composition, electronic structure, and surface properties of materials with high sensitivity and precision. In the context of 2D perovskites, XPS offers unique insights into their thermal stability by elucidating the evolution of chemical bonds and surface species upon heating. Our study examined the thermal stability of mechanically exfoliated 2D $(BA)_2PbI_4$ and $(PEA)_2PbI_4$ perovskite flakes by heating them in air followed by XPS measurements. The figure 5 depicts the XPS spectra of core-levels including Pb 4f, I 3d, and N1s, obtained from freshly exfoliated flakes and those subjected to heat treatment. Through the analysis and comparison of these core-level spectra between freshly exfoliated and heat-treated samples, we can discern the chemical alterations experienced by each element because of degradation induced by heating in air. Pb 4f (Figure 5a,b) and I 3d (Figure 5c,d) core-level spectra of freshly exfoliated flakes of both perovskites are dominated by $Pb^{+2}$ and $I^{-1}$ spin-orbit doublets with a Pb $4f_{7/2}$ component located at 138.05 eV (Pb $4f_{5/2}$ at 142.93 eV) and I $3d_{5/2}$ component at 618.9 eV (I $3d_{3/2}$ at 630.36 eV), respectively. These components are attributable to $Pb^{+2}$ and $I^{-1}$ atomic states in the inorganic octahedral of the $[PbI_6]^{4-}$ units of the perovskite lattice. One singlet representing C-N bonds in organic cations $PEA^+$ in $PEA_2PbI_4$ located at 401.6 eV (Figure 5e) and $BA^+$ in $BA_2PbI_4$ located at 401.7 eV (Figure 5f) and, were detected in N 1s core-levels spectra. The spectra of both fresh samples don't show any evidence of phase segregation indicating a good quality of the synthesised crystals (see Materials and Methods).

Upon heating, two more spin-orbit doublets are observed in the Pb 4f spectra of both perovskites. Here, the Pb $4f_{7/2}$/Pb $4f_{5/2}$ peaks at 136.33/141.1 eV and 138.41/143.29 eV are attributed to metallic lead ($Pb^0$) and $PbI_2$ respectively. The presence of $Pb^0$and $PbI_2$ indicates the degradation of the $[PbI_6]^{4-}$ octahedral units in the crystal structure. This is further confirmed by the measurement of I 3d spectra which shows a gradual decrease in the intensity of peaks with an increase in temperature (figure 5c, d). Temperature-induced degradation of both 2DPVK is primarily detected by a decrease in the intensity of the N 1s peak that represents the C-N bonding in $(PEA)^+$ and $(BA)^+$ organic cations (figure 5e,f), indicating the thorough elimination of organic cations from the perovskite lattice. The N 1 s spectra also confirm the existence of nitrogen defects in the perovskite lattice located at 399.6 eV. Interestingly, the maximum amount of defects are observed at 80°C in both $(BA)_2PbI_4$ and $(PEA)_2PbI_4$ perovskites. Sample heating induces ion's migration resulting a significant drop of the I/Pb and N/Pb atomic ratios (see Figure S6 in Supplementary material) and creation of several vacancies and defects in the crystal structure. These findings unequivocally validate the instability of the 2DPVK at elevated temperatures. Compared to our previous XPS results obtained for 2DPVKs slowly aged at ambient conditions [6], the degradation mechanism induced by thermal heating does not lead to the production of lead salts on the sample surface, as salts cannot be formed without a presence of moisture (water is evaporated by heating), which is a requisite reactant for the reaction [67]. For a deeper understanding, we performed quantitative estimations of the Pb 4f atomic ratios of individual components $Pb^0$, $Pb^{+2}$ and $PbI^2$ for $PEA_2PbI_4$ (Figure 5g) and $BA_2PbI_4$ (Figure 5h) flakes as function of thermal heating in air. Comparing their temperature evolution, the main difference between both samples can be found in $PbI_2$ and $Pb^0$ ratios (see equations 1 and 2). The amount of $PbI_2$ depends on the perovskite degradation rate (equation 1) and its further decomposition into $Pb^0$ (equation 2), affected by material and surrounding conditions [61]. $PEA_2PbI_4$ degrades preferably in $PbI_2$ accompanied by the evaporation of volatile compounds (equation 1). Whereas $BA_2PbI_4$ degradation mechanisms are more destructive, involving faster evaporation and phase segregation leading to a higher amount of $Pb^0$ than $PbI_2$ as a degradation products (equation 2). These results are in good agreement with higher photostability and thermal stability of $PEA_2PbI_4$ observed in Raman analysis.

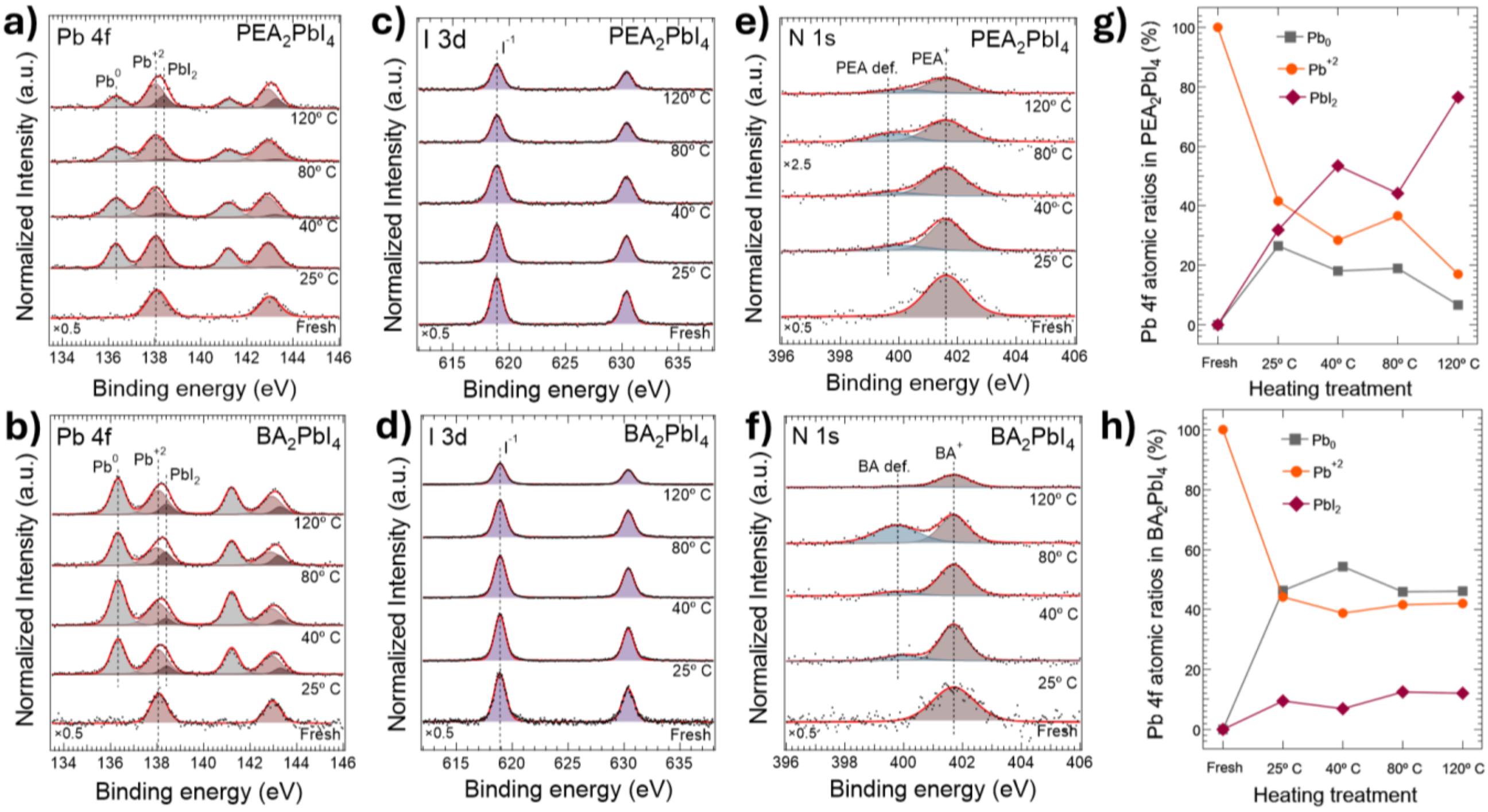

***Figure 5.*** *High resolution X-ray photon emission spectroscopy (XPS) analysis recorded immediately after exfoliation of fresh $PEA_2PbI_4$ (top panel) and $BA_2PbI_4$ (bottom panel) flakes and after their heating treatment (25-120 $^oC$) under ambient air conditions. XPS comparison of freshly exfoliated and heat treated $PEA_2PbI_4$ flakes with corresponding (a) Pb 4f, (c) I 3d and (d) N 1s core levels (top panel). XPS comparison of freshly exfoliated and heat treated $BA_2PbI_4$ flakes with corresponding (b) Pb 4f, (d) I 3d and (f) N 1s core levels (bottom panel). Pb 4f atomic ratios of individual components depicted in (a,b) for (g) $PEA_2PbI_4$ and (h) $BA_2PbI_4$, flakes as function of thermal heating in air.*

## 3 Conclusions

In this work, we systematically examined the photo- and thermal-stability of mechanically exfoliated $(BA)_2PbI_4$ and $(PEA)_2PbI_4$ lead halide 2DPVKs nanoflakes under ambient air conditions. Even though freshly exfoliated $(PEA)_2PbI_4$ and $(BA)_2PbI_4$ nanoflakes initially exhibit similar optical properties at room temperature, the type of organic spacer molecule significantly affects their environmental stability. Both crystal phases degraded similarly, resulting in the removal of organic components and fast crystal phase decomposition into $PbI_2$ and $Pb^0$. However, $PEA_2PbI_4$ demonstrated greater resistance to high-power laser illumination and thermal decomposition under ambient air conditions than $(BA)_2PbI_4$. $PEA_2PbI_4$ degrades preferably into $PbI_2$ accompanied by evaporation of volatile compounds from the lattice and to a lesser extent into $Pb^0$ formation. In contrast, the degradation mechanisms of $BA_2PbI_4$ lead to faster evaporation of volatile compounds, $PbI_2$ phase decomposition into $Pb^0$, and crystal destruction.

We demonstrate that $PEA^+$ organic cation may serve as a more effective passivation agent, promoting improved crystallinity and reducing defect trap formation within the lead halide 2DPVKs lattice. Thus, the use of this organic cation might enhance both device stability and performance. We conclude that structural engineering of 2DPVK materials, combined with surface passivation or encapsulation strategies, is essential for their future applications.

## 4. Materials and Methods

### 4.1 Synthesis and sample preparation

Text adapted from our paper about Donor-Acceptor pairs[33]

Materials were prepared by modifying a synthetic procedure that has been reported elsewhere [75]. For synthesizing $PEA_2PbI_4$ (n = 1), PbO powder (1116 mg, 5 mmol) was dissolved in a mixture of 57% w/w aqueous HI solution (5.0 mL, 38 mmol) and 50 % aqueous $H_3PO_2$ (0.85 mL, 7.75 mmol) by heating to boiling under vigorous stirring for about 10 min until a formation of a bright yellow solution A was observed. In a separate beaker, $C_6H_5CH_2CH_2NH_2$ (phenetylamine, PEA) (628 $\mu$L, 5 mmol) was neutralized with HI 57% w/w (2.5 mL, 19 mmol) in an ice bath, resulting in a clear pale-yellow solution B. The solution B was slightly heated until a solid precipitate or the formation of a suspension has dissolved. Once both solutions have been prepared, we proceed to the slow addition of Solution B to Solution A. A quick addition could produce a precipitate, which was subsequently dissolved as the combined solution was heated to boiling. After keeping the solution at boiling point for 10 min, the stirring was then discontinued and the solution was left to cool to room temperature, promoting the crystalization oforange rectangular-shaped plates along the time. The crystallization was deemed to be complete after 1 h.. The crystals were isolated by suction filtration and washed with cold diethyl ether.

Text adapted from our paper based on2D perovskite's long time degradation [16]

For the preparation of $BA_2PbI_4$ (n = 1), 0.5 g (2.24 mmol) of PbO powder was dissolved in stabilized HI (5.0 mL, 38 mmol) under constant magnetic stirring, and heating to boiling for about 5 min until a bright yellow solution was observed. In a separate beaker, 0.43 mL (4.48 mmol) of n-butylamine (n-$CH_3(CH_2)_3NH_2$) was neutralized with HI 57% w/w (2.5 mL, 19 mmol) in an ice bath resulting in a clear pale-yellow solution. The addition of the n-$CH_3(CH_2)_3NH_3I$ solution to the $PbI_2$ solution produces a precipitate, which was subsequently dissolved under heating the combined solution to boiling. After stop the heating/stirring process (5 minutes after obtaining a homogeneous solution, the solution was cooled down to room temperature slowly (5° C/h), obtaining orange crystal sheets. The crystallization was deemed to be complete after a night. The crystals were isolated by suction filtration, washed with cold diethyl ether and thoroughly dried under reduced pressure.

Flakes of 2D lead halide perovskites (2DPVKs) were mechanically exfoliated from the synthetized well-defined $BA_2PbI_4$ (*n*=1 phase) and $PEA_2PbI_4$ (*n*=1 phase) single crystals onto clean $SiO_2$ (270 nm)/Si (500 µm) or N-Type conductive Si substrates with dimensions of 1x1 cm using a scotch tape before photo- and thermal- stability investigations.

### 4.2 X-ray diffraction analysis

X-ray diffraction (XRD) measurements of crystals were taken by an X-ray diffractometer (D8 Advance, Bruker-AXS) (Cu Kα, wavelength λ = 1.5406 Å) from 3 to 60° ($BA_2PbI_4$) and 5 to 60° ($PEA_2PbI_4$). The theoretical powder X-ray diffraction (PXRD) patterns were obtained via computational simulation using the VESTA software package. The simulations were generated utilizing the lattice parameters and atomic coordinates derived from the single-crystal X-ray diffraction (SCXRD) data reported in the cited literature [36,37]. This provided an ideal reference model for the target crystalline phases, enabling a direct comparison with the experimental diffractograms to confirm structural phase purity.

### 4.3 Raman spectroscopy

Micro-Raman spectroscopy measurements were carried out using a confocal Horiba-MTB Xplora Raman spectrometer to study the structural and chemical properties of freshly exfoliated and photo and thermally degraded $(PEA)_2PbI_4$ and $(BA)_2PbI_4$ samples under ambient air. All Raman measurements were obtained using a 50x objective with numerical aperture (N.A.) = 0.5 and working distance (WD) = 1.3 mm. Room-temperature Raman spectra were recorded using infrared laser excitation (λ = 785 nm) with a laser power of 6.4 mW, an illumination time of 90 s, and 600 lines/mm diffraction grating. The scattered light was measured with a Peltier-cooled CCD detector. These conditions minimize laser induced sample degradation.

For photodegradation studies, freshly exfoliated samples were irradiated by a high-energy green laser (λ = 532 nm) with laser spot diameter diameter ≈3.4 µm using low (68 µW) and high (4.5 mW) laser power and various irradiation times (60 s, 90 s and 120 s). To examine sample degradation, Micro-Raman spectra were recorded immediately after the photothermal treatment at the same sample location. To investigate PL degradation, we used the same micro-Raman spectroscopy setup with, illuminating for a short time and low power with the green laser beam (λ =532 nm, 7.4 µW, 0.15 s) to minimize uncontrolled sample degradation. PL spectra were taken at the same sample location where the photodegradation treatment was applied (for photodegradation, the green, λ =532 nm laser was used during for 120 s). Low (68 µW) and high (4.5 mW) laser powers were used at different spots of the samples. At this high photon energy (λ =532 nm) the PL signal prevails over the Raman signal.

Temperature-dependent micro-Raman spectra were obtained by coupling a Linkam TS1500 heating stage to the confocal Raman microscope. The heating rate was set to 5 °C $min^{-1}$. To ensure thermal equilibrium at each target temperature, Raman spectra were recorded five minutes after the stage reached the selected temperature step (25°C, 40°C, 80°C and 120°C) by using a lower energy red laser (λ = 785 nm, 6.4 mW, 90) minimizing laser-induced degradation processes . The stability of the materials was monitored through the attenuation of the signal at 130 $cm^{-1}$, which originates from the organic salts BA and PEA. We also identified a stronger contribution from the A2u signal, indicating the dominant presence of $PbI_2$. Notably, the two-dimensional material fabricated using PEA exhibits considerable stability even at temperatures of 80 °C, in contrast to the material made with BA, which undergoes rapid degradation when heated—even below 80 °C.

### 4.4 High-resolution X-ray photoelectron spectroscopy

High energy resolution X-ray photoelectron spectroscopy (XPS) measurements were performed in a SPECS GmbH system (base pressure $1.0 \times 10^{-10}$ mbar) equipped with a PHOIBOS 150 2D-CMOS hemispherical analyzer. Photoelectrons were excited with the Al-$K_a$ line (1486.7 eV) of a monochromatic x-ray source µ-FOCUS 500 (SPECS GmbH). Measurements were taken at room temperature with a pass-energy of 20 eV. For all of the HR-XPS measurements, freshly exfoliated (25°C) and previously thermal heated samples in air (heating temperature of 40°C, 80°C and 120°C) were exposed to air and placed in ultra-high vacuum (UHV) conditions (with a base pressure of analysis chamber lower than $1.0\times10^{-10}$ mbar).

**5. Acknowledgements**

The present work has been supported by the projects MFA2022/009 (SPINO2D), CIPROM/2021/078, Prometeo/2021/082, CIPROM-2021-75, and CIPROM/2024/04 of the Generalitat Valenciana; as well as Projects PID2020-112507GB-I00, PID2023-146181OB-I00, PID2024-161277OB-I00 and PID2022-140090OB-C21/AEI/10.13039/501100011033/FEDER of the State Investigation Agency and the TED2021-132656B-C22 granted by the Call 2021 - "Ecological Transition and Digital Transition Projects" promoted by the Ministry of Science and Innovation, funded by the European Union within the "NextGeneration" EU program, the Recovery, Transformation, and Resilience Plan. M. E. C. acknowledges financial support from the Santiago Grisolia Program of the Generalitat Valencia (grant CIGRIS/2021/150). A.B and T.B acknowledge financial support from the University of Valencia (through the Program "Universitat de València with Ukraine", the Generalitat Valenciana through the Program "Acoge CV-UCRANIA personal investigador" of the Conselleria de Innovación, Universidades, Ciencia y Sociedad Digital, and the Ministry of Universidades through the program "Plan de Acción Universidad-Refugio".

Raman and photoluminescence measurements were carried out at the Atomic and Molecular Spectroscopy Unit of the Central Service for Experimental Research of the University of València (SCSIE).

**Supporting information**

Figure S1: (a, b) AFM measurements of two different $PEA_2PbI_4$ flakes. Figure S2: Scanning Electron Microscopy (SEM) images of exfoliated $PEA_2PbI_4$ flakes. Figure S3. Photoluminescence statistic of several exfoliated $PEA_2PbI_4$ and $BA_2PbI_4$ flakes as function of the peak emission central energy and Full Width at Half Maximum (FWHM). Figure S4: The Optical images comparison of freshly exfoliated (a,b) $PEA_2PbI_4$ and (c,d) $BA_2PbI_4$ samples aging in air over the period of 3 weeks with absence and presence of ambient light exposition showing high impact of light on the crystal structure degradation. Figure S5: The photoluminescence background in (a) $PEA_2PbI_4$ and (b) $BA_2PbI_4$ flakes was recorded with the micro-Raman spectrometer by using green excitation but with extremely low power and laser excitation time (532 nm excation wavelenth, 7.4 µW, 0.15 s). This should minimize sample degradation. Photoluminescence spectra obtained before and after the photothermal treatment at low (68 µW) and high (4.5 mW) laser powers are shown. Figure S6: High resolution X-ray photon emission spectroscopy (XPS) analysis of freshly exfoliated and heat treated $PEA_2PbI_4$ and $BA_2PbI_4$ flakes. Comparison of (a) I/Pb and (b) N/Pb atomic ratios for $PEA_2PbI_4$ and $BA_2PbI_4$ flakes as function of thermal heating in air, considering core-level cross sections at the used photon energy.